\documentclass[aps,twocolumn,prl,superscriptaddress]{revtex4-2}
\usepackage{graphicx,color}
\usepackage{amsmath}
\usepackage{amssymb}
\usepackage{amsthm}
\usepackage{hyperref}
\usepackage{enumitem}
\usepackage{bm}

\newcommand{\tr}{\mathop{\mathrm{tr}}\nolimits}

\newcommand{\diag}{\mathop{\mathrm{diag}}\nolimits}
\def\d{\mathrm{d}}

\newcommand{\rmi}{\mathrm{i}}
\newcommand{\rme}{\mathrm{e}}
\newcommand{\ket}[1]{|{#1}\rangle}
\newcommand{\bra}[1]{\langle{#1}|}

\definecolor{dgreen}{rgb}{0,0.5,0}

\definecolor{dblue}{rgb}{0,0,0.6}

\definecolor{dred}{rgb}{0.784,0,0}

\definecolor{delete}{cmyk}{0.5,0,0,0}

\begin{document}
\title{KAM-Stability for Conserved Quantities in Finite-Dimensional Quantum Systems}
\author{Daniel Burgarth} 
\affiliation{Center for Engineered Quantum Systems, Dept.\ of Physics \& Astronomy, Macquarie University, 2109 NSW, Australia}
\author{Paolo Facchi}
\affiliation{Dipartimento di Fisica and MECENAS, Universit\`a di Bari, I-70126 Bari, Italy}
\affiliation{INFN, Sezione di Bari, I-70126 Bari, Italy}
\author{Hiromichi Nakazato}
\affiliation{Department of Physics, Waseda University, Tokyo 169-8555, Japan}
\author{Saverio Pascazio}
\affiliation{Dipartimento di Fisica and MECENAS, Universit\`a di Bari, I-70126 Bari, Italy}
\affiliation{INFN, Sezione di Bari, I-70126 Bari, Italy}
\author{Kazuya Yuasa}
\affiliation{Department of Physics, Waseda University, Tokyo 169-8555, Japan}
\date{\today}
\begin{abstract}
We show that for any finite-dimensional quantum systems the conserved quantities can be characterized by their robustness to small perturbations: for fragile symmetries small perturbations can lead to large deviations over long times, while for robust symmetries their expectation values remain close to their initial values for all times. This is in analogy with the celebrated Kolmogorov-Arnold-Moser (KAM) theorem in classical mechanics. To prove this remarkable result, we introduce a resummation of a perturbation series, which generalizes the Hamiltonian of the quantum Zeno dynamics.
\end{abstract}
\maketitle

Symmetries and conserved quantities are the cornerstones of modern theoretical physics \cite{Gross}. In quantum mechanics, it is well known that conserved quantities are characterized by observables which commute with the system Hamiltonian. Here, we show that this characterization is incomplete, because some symmetries in quantum mechanics are more conserved than others.

More precisely, we can consider the \emph{robustness} of symmetries. Some fundamental symmetries (such as those related to superselection rules \cite{Moretti}) are considered almost unbreakable in nonrelativistic quantum mechanics, while other, accidental \cite{Alhashimi}, symmetries are easily perturbed.

We introduce such distinction into fundamental, robust symmetries and accidental, fragile ones in an analogous, but much more applied context, namely one provided by a time-independent Hamiltonian on a finite-dimensional quantum system. This Hamiltonian $H$ acts as a reference, with respect to which we will introduce a unique decomposition of any symmetry $S$ satisfying $[H,S]=0$ as
\begin{equation}
S=S_{\textrm{robust}}+S_{\textrm{fragile}}.
\end{equation} 
We will show that the \emph{robust} component of $S$ remains almost conserved [up to a term $O(\varepsilon)$] for \emph{all} times and for \emph{any} small time-independent perturbation $\varepsilon V$, while for the \emph{fragile} component most perturbations will accumulate large amounts of change over time. As an alternative view, for any robust symmetry there is a slightly modified observable which is conserved in the perturbed system, $S_{\textrm{robust}}\rightarrow S_{\textrm{robust}}^\varepsilon$, while for fragile symmetries there is not. Such conserved quantities were constructed recently in many-body systems for specific perturbations \cite{Caux}, while we provide a general construction.

The importance of robust observables is exemplified by analogue quantum simulations \cite{Nori}, where the aim is to run a complex Hamiltonian long enough such that observable quantities are no longer easily computable by classical computers. The problem is, however, that small perturbations in the lab are not under control and can destroy the reliability of the simulation \cite{Sarovar,Schwenk}. On the other hand, as we show below, the expectation values of robust observables remain reliable even on the long term.

More fundamentally, our result is in close analogy to the KAM perturbation theory in classical mechanics \cite{ref:ThirringClassicalPhysics,ref:ArnoldBook}, which proved the long-term stability of planetary orbits, despite accumulating perturbations. Quantum mechanical versions of KAM perturbation have been considered previously by Scherer \cite{ref:Scherer95} to mimic a superconvergent series. Our focus, instead, is  an algebraic approach based on the adiabatic theorem, enabling us to provide nonperturbative bounds and generalizations to open systems. This way, we prove the KAM-stability in finite-dimensional quantum mechanics.

How can we characterize which observables are fragile, and which are robust? Under which conditions are there robust ones, and just \emph{how} robust are they?
In the unperturbed system, the conserved quantities are the observables commuting with $H$, given by all Hermitian matrices which are block-diagonal with respect to the eigenspaces of $H$. They may share the degenerate eigenspaces of $H$ or they may lift their degeneracy. In this Letter, we will show that this precisely distinguishes robust and fragile symmetries. Moreover, unless $H$ is the identity, there always exist nontrivial robust symmetries.

\emph{Fragile symmetries.---}%
First, consider a symmetry $M$ which breaks degeneracy in an eigenspace of $H$. 
We show that such a conserved quantity is not robust against perturbation.
For instance, take two simultaneous eigenstates of $H$ and $M$, say $\ket{e_1}$ and $\ket{e_2}$, belonging to the same eigenspace of $H$ but belonging to different eigenspaces of $M$, i.e., $H\ket{e_1}=e \ket{e_1}$ and $H \ket{e_2}=e\ket{e_2}$, while $M\ket{e_1}=m_1\ket{e_1}$ and $M\ket{e_2}=m_2\ket{e_2}$, with $\Delta = m_1 - m_2>0$. Let us take $V=|e_1\rangle\langle e_2|+|e_2\rangle \langle e_1|$ as a perturbation and consider $H+\varepsilon V$. If we focus on initial states $\ket{\psi}$ in the subspace spanned by $\{\ket{e_1},\ket{e_2}\}$, the problem is reduced to a two-dimensional problem.
Take for instance $\ket{e_1}$ as an initial state.
We find
\begin{equation}
\langle M\rangle^\varepsilon_t-\langle M\rangle_t = \langle M\rangle^\varepsilon_t-\langle M\rangle_0
=-\Delta\sin^2\varepsilon t ,
\end{equation}
where the expectations $\langle{} \cdot{} \rangle_t$  and $\langle {}\cdot{} \rangle_t^\varepsilon$  are taken with respect to states evolved under the free  and the perturbed evolution, $\rme^{-\rmi t H}\ket{\psi}$ and $\rme^{-\rmi t (H+\varepsilon V)}\ket{\psi}$, respectively.
At time $t=\pi/(2\varepsilon)$ the error is $\Delta$, which is independent of $\varepsilon$. This kind of example can be constructed for any $M$ which is nondegenerate within a subspace of $H$, and we conclude that such conserved observables are \emph{fragile}.

\emph{Robust symmetries.---}%
Second, consider a conserved observable which acts uniformly within each eigenspace of $H$. 
We may write $M=\sum m_kP_k$, where $\{P_k\}$ are the spectral projections of $H=\sum_ke_kP_k$ (with $e_k\neq e_{\ell}$ for $k\neq \ell$ and $P_k P_\ell =\delta_{k\ell}P_\ell$).
Using results on the quantum Zeno dynamics~\cite{ref:QZS,ref:PaoloSaverio-QZEreview-JPA,ref:unity1,HamazakiPRL2020}, one can show that such observables are endowed with some intrinsic robustness with respect to small perturbations $\varepsilon V$, with $\|V\|=1$. Indeed, we have a bound~\cite{note:KAM_SM}
  \begin{equation}
\delta_{\mathrm{Z}}(t) = \|
\rme^{\rmi t(H+\varepsilon V)}
-
\rme^{\rmi t (H+\varepsilon V_{\mathrm{Z}})}
\|
\leq\frac{2 \sqrt{d}}{\eta}\varepsilon(1+\varepsilon t),
\label{eq:deltaZ}
\end{equation}
where 
 $d$ is the number of distinct eigenvalues of the Hamiltonian $H$, 
$\eta=\min_{k\neq\ell}|e_k-e_\ell|$  is the spectral gap of $H$, 
and  
$V_{\mathrm{Z}}=\sum_kP_kV P_k$ is the \emph{Zeno Hamiltonian}~\cite{ref:QZS,ref:PaoloSaverio-QZEreview-JPA}. 
By construction $[M,V_{\mathrm{Z}}]=0$, and we obtain~\cite{HamazakiPRL2020,note:KAM_SM}
 \begin{equation}
\| M_t^\varepsilon- M \| 
\leq 2 \|M\| \delta_{\mathrm{Z}}(t) 
\leq \frac{4 \sqrt{d}}{\eta}\|M\|\varepsilon (1+\varepsilon t),
\end{equation}
where $M_t^\varepsilon= \rme^{\rmi t(H+\varepsilon V)} M \rme^{-\rmi t(H+\varepsilon V)}$ is the perturbed evolution of  observable $M$.
This bound however is informational as far as it is less than the  trivial bound $2\|M\|$, which is not for sufficiently large times $t$.

This is, anyway, just an upper bound, and it might be a loose bound. 
Let us look more carefully at a two-dimensional example again, 
and show that there are indeed perturbations $V$ such that $\delta_{\mathrm{Z}}(t)$ in~\eqref{eq:deltaZ} saturates the trivial bound 2, for every $\varepsilon$ however small. Consider $H=\sigma_z$ and $V=\sigma_x$, the third and first Pauli matrices, respectively.
In this case, we have $V_{\mathrm{Z}}=0$, and we compute $\delta_{\mathrm{Z}}(t)=\|\rme^{\rmi t(\sigma_z+\varepsilon \sigma_x)}-\rme^{\rmi t \sigma_z}\|$.
This in general is a complicated quasiperiodic function, and one has $\sup_{t} \delta_{\mathrm{Z}}(t) =2$. Actually,
one can choose a specific sequence $\varepsilon_n =\frac{\sqrt{4 n+3}}{2 n+1}\to0$ such that at times $t_n=\frac{\pi}{\sqrt{1+\varepsilon_n^2}-1}$ the norm distance saturates, $\delta_{\mathrm{Z}}(t_n)=2$.
This shows that in general the Zeno Hamiltonian $V_{\mathrm{Z}}$ is not a good approximation for long times.

However, notwithstanding the negative result about the smallness of the distance $\delta_{\mathrm{Z}}(t)$ in~\eqref{eq:deltaZ}, the conserved quantity $M=\sum m_kP_k$ considered above \emph{is actually stable} for all times, \emph{eternally}.
The key idea behind the above surprising phenomenon is to choose an ($\varepsilon$-dependent) approximation of $V$ which has the same block structure as $H$ and is therefore commutative with $M$. 
The Zeno Hamiltonian $V_{\mathrm{Z}}$ is not a good choice.
To make the point, consider again the above two-dimensional example with $H=\sigma_z$ and $V=\sigma_x$, and now choose, in place of $V_{\mathrm{Z}}$, the operator
$V_{H}(\varepsilon)
= \varepsilon^{-1} (\sqrt{1+\varepsilon^2}-1)\sigma_z$
as an approximation of $V$.
Obviously, $[V_H,H]=0$. Moreover, 
$V_{H}(\varepsilon) =  V_{\mathrm{Z}}+ O(\varepsilon)$.
With this choice, we get 
\begin{align}
\delta(t)
&=
\|
\rme^{\rmi t(\sigma_z+\varepsilon\sigma_x)}
-
\rme^{\rmi t [\sigma_z+\varepsilon V_H(\varepsilon)]}\|
\nonumber\\
&=\sqrt{2\left(1-\frac{1}{\sqrt{1+\varepsilon^2}}\right)}\left|\sin(t \sqrt{1/\varepsilon^2+1})\right|
\le\varepsilon.
\end{align}
Remarkably, this bound is independent of time $t$, and implies that~\cite{note:KAM_SM}
\begin{equation}
\| M^\varepsilon_t- M \| 
\leq 2 \|M\|  \delta(t)
\leq 2 \|M\| \varepsilon ,
\end{equation}
\emph{for all times} and 
for any observable of the form 
$M = \diag(m_1,m_2)$.
Such observables are the \emph{robust} observables.

\emph{General result.---}%
Of course, we did not just guess $V_H(\varepsilon)$ arbitrarily.
We discovered a way of constructing such \emph{eternal} block-diagonal approximations for any finite-dimensional quantum systems, including noisy systems with Lindbladians. They can be seen as resummation of  a perturbative series, whose zeroth-order term is the Zeno Hamiltonian $V_H(0)= V_{\mathrm{Z}}$.
Its theory, its proof, and generalizations are discussed in great detail in Ref.~\cite{eternal}.

The crucial ingredient is that the block-diagonal approximation $H+ \varepsilon V_H(\varepsilon)$, unlike $H+ \varepsilon V_{\mathrm{Z}}$, can be chosen to have the \emph{same spectrum} of $H+\varepsilon V$, and  thus to be unitarily equivalent to it: $H+ \varepsilon V_H(\varepsilon) = W_\varepsilon^\dagger (H+\varepsilon V) W_\varepsilon$, with a unitary $W_\varepsilon = \openone + O(\varepsilon)$~\cite{note:KAM_SM}.
This is a necessary condition since geometrically the evolution of a Hamiltonian with $d$ distinct eigenvalues yields a (quasi-)periodic motion of a point on a torus. Two motions with different frequencies, however small the differences may be, will eventually accumulate a divergence of $O(1)$. The only way to avoid this slow drift is that the two motions be isochronous, that is the two Hamiltonians be isospectral. In such a case we get
\begin{align}
\delta_\infty 
&= \sup_t \| \rme^{\rmi t(H+\varepsilon V)} - \rme^{\rmi t [H+ \varepsilon V_H(\varepsilon)]} \| 
\nonumber\\
&= \sup_t \| \rme^{\rmi t(H+\varepsilon V)} - W_\varepsilon^\dagger\rme^{\rmi t (H+ \varepsilon V)}W_\varepsilon \| 
< \frac{7\sqrt{d}}{\eta} \varepsilon
\end{align}
(see Refs.~\cite{note:KAM_SM,eternal} for its explicit bound).
It follows that \emph{any} quantum system has robust conserved ($M_t = M$) quantities, such that for every perturbation $\varepsilon V$, 
\begin{equation}
\| M^\varepsilon_t- M \|
\leq 2 \|M\| \delta_\infty
= O(\varepsilon) ,
\label{eq:supt}
\end{equation}
for all times \cite{note:KAM_SM},
where 
$M^\varepsilon_t = \rme^{\rmi t (H + \varepsilon V)} M \rme^{-\rmi t (H + \varepsilon V)}$,
and they are precisely given by 
\begin{equation}
M=\sum_k m_k P_k,
\end{equation} 
with $H=\sum_ke_kP_k$. 
All other conserved quantities are fragile, as the distance~\eqref{eq:supt} becomes $O(1)$.

While this is a complete characterization of robust conserved quantities, the representation in terms of spectral projections requires diagonalization of the Hamiltonian $H$ and is impractical for high-dimensional systems.
However, given $H^n=\sum_ke_k^nP_k$, one can invoke the invertibility of the Vandermonde matrix $(e_k^{j-1})$ 
to see that $M=\sum_{n=0}^{d-1}c_nH^n$,  where $d$ is the number of distinct eigenvalues of the Hamiltonian $H$.
This means that any primary matrix function $f(H)$ of the Hamiltonian $H$ is robust.
If the original Hamiltonian is sparse, for instance low-order polynomials can be constructed efficiently.
In particular, we obtain that for any state the energy expectation value and the variance,
\begin{equation}
\langle H\rangle^\varepsilon_t- \langle H\rangle_0= O(\varepsilon) ,
\quad \langle\Delta H^2\rangle^\varepsilon_t-\langle \Delta H^2\rangle_0=O(\varepsilon),
\end{equation}
remain close to their unperturbed values \emph{forever}.
This is easily generalized to higher moments.

We can also rephrase the fact that any robust observable is a polynomial function of $H$ in terms of the symmetries of $H$. That is, $M$ is robust if and only if it shares \emph{all} symmetries of $H$: for any $C$ such that $[H,C]=0$ we also have $[M,C]=0$~\cite{ref:MatrixFunctions-Higham}.

Finally, let us emphasize that the above characterization provides a natural decomposition of any observable $M$ into three parts: one dynamical part which is not conserved by $H$, one which is conserved but is fragile to perturbations, and one which is  robust. 
The nonconserved part is off-diagonal with respect to the spectral projections of~$H$, 
\begin{equation}
\label{off}
M_\text{noncons}= M- M_\text{cons} = M-\sum_k P_kM P_k,
\end{equation}
the robust component is the diagonal component which acts trivially within the eigenspaces of $H$,
\begin{equation}
\label{robust}
M_\text{robust}=\sum_k \frac{1}{d_k}\tr(P_kMP_k)P_k,
\end{equation}
with $d_k$ being the dimension of the $k$th eigenspace, and the fragile part is the remaining diagonal part,
\begin{equation}
\label{fragile}
M_\text{fragile}= 
\sum_k P_k \left(M - \frac{1}{d_k}\tr(P_kMP_k)\right) P_k.
\end{equation}

\begin{figure}
\includegraphics[width=0.45\textwidth]{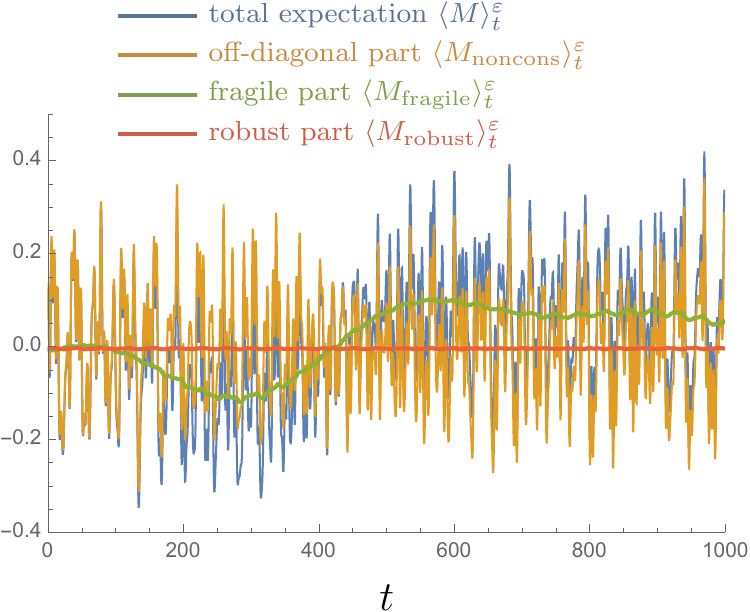}
\caption{Dynamics of the expectation of a randomly picked observable $M$ in a Heisenberg chain of $N=4$ as a function of time. We show its decomposition into nonconserved, robust, and fragile parts [Eqs.~(\ref{off})--(\ref{fragile})]. The initial state and the perturbation $V$ are chosen randomly and the perturbation strength $\varepsilon=0.02$, where $H$, $V$, $M$ are normalized to $\|H\|=\|V\|=\|M\|=1$. Shown is one realization.}
\label{fig:random}
\end{figure}
\begin{figure}
\includegraphics[width=0.45\textwidth]{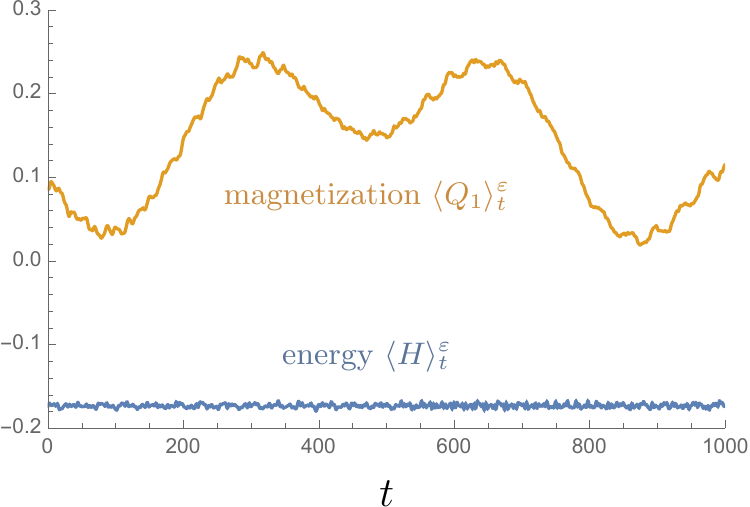}
\caption{Same setup and realization as in Fig.~\ref{fig:random}, but now showing the dynamics of the expectation of $H$ (robust) and $Q_1=\sum_{n=1}^N \sigma_{n,z}$ (fragile), where $Q_1$ is also normalized to $\|Q_1\|=1$.}
\label{fig:physical}
\end{figure}
\emph{Integrable example.---}%
While an unambiguous and universal definition of integrability for quantum systems is still lacking~\cite{ref:Caux-QuantumIntegrability,ref:Braak-RabiExact}, we take the Heisenberg chain as a typical example of a system which we think of as integrable. 
The Hamiltonian acts on $N$ qubits and is given by $H=-J\sum_{n=1}^{N}\bm{\sigma}_n\cdot\bm{\sigma}_{n+1}$
where $\bm{\sigma}_n =(\sigma_{n,x}, \sigma_{n,y}, \sigma_{n,z})$ is the vector of Pauli matrices acting on the $n$th qubit, and we impose the periodic boundary conditions $\bm{\sigma}_{N+1}=\bm{\sigma}_{1}$.
The Heisenberg chain can be solved analytically by the algebraic Bethe ansatz. The corresponding conserved charges $Q_2,\dots, Q_n$ can be generated using the boost operator $B=\frac{1}{2}\sum_{n=1}^{N}n\bm{\sigma}_n\cdot \bm{\sigma}_{n+1}$ as $Q_{n+1}=-\rmi [B,Q_n]$ with $Q_2=H$, and $Q_n$ acts nontrivially on sets of $n$ neighbors on the chain only~\cite{Grabowski}. Combined with the total magnetization $Q_1=\sum_{n=1}^N \sigma_{n,z}$, they provide a maximal Abelian algebra. These conserved charges are the pinnacle of integrability. However, perhaps counterintuitively, they are \emph{fragile}: because the powers of $H$ do include long-range interactions, it is easily seen that none of the charges are robust. Incidentally, this shows that the findings in Ref.~\cite{Caux} are restricted to specific perturbation classes. A simple example is given by the total magnetization $Q_1$: due to the rotational invariance of $H$, we could have equally chosen the magnetization in another direction, say $\tilde{Q}_1= \sum_{n=1}^N \sigma_{n,x}$. As a perturbation however, $\tilde{Q}_1$ causes the expectation value of $Q_1$ to oscillate and deviate vastly from its original value.
For instance, if we start with a $z$-polarized state, we obtain $\langle Q_1\rangle_t^\varepsilon=\cos (\omega t) N $ for some $\omega$ depending on the perturbation strength $\varepsilon$. We show numerical examples of the evolution of a randomly chosen observable (Fig.~\ref{fig:random}) as well as physical  ones (Fig.~\ref{fig:physical}).

\emph{Thermalization.---}%
It is one of the most celebrated results in mathematical quantum statistical mechanics that the KMS state [the Gibbs state $\propto\exp(-\beta H)$] is the unique state which maximizes entropy, stationary under the time evolution of the Hamiltonian $H$, and \emph{robust under perturbations}~\cite{Haag}. However, till date this was only considered for short times. The remarkable consequence of our characterization of robust observables implies that 
\begin{equation}
\rme^{-\rmi t(H+\varepsilon V)}	  \exp(-\beta H) \rme^{\rmi t(H+\varepsilon V)}	 =  \exp(-\beta H ) + O(\varepsilon),
\end{equation}
uniformly in time for any finite-dimensional system.

Again, perhaps surprisingly, generalized Gibbs ensembles \cite{ref:PolkovnikovRMP,ref:EisertFriesdorfGogolin-ReviewNatPhys,ref:GooldReviewJPA,ref:GogolinEisertReview,ref:VidmarRigolGGEreview,ref:CalabresePhysicaA,ref:MoriIkedaKaminishiUeda-ReviewJPB} such as $\exp(-\sum_j  \beta_j Q_j)$ for integrable charges are \emph{not} robust.

\emph{Open systems.---}%
How do we generalize this to Lindbladian dynamics? For a Lindbladian $\mathcal{L}$, it would be natural to consider $\mathcal{M}=\sum_k m_k \mathcal{P}_k$, with $\{\mathcal{P}_k\}$ the spectral projections of $\mathcal{L}$, as a candidate for a robust symmetry. However, it is easy to see that the trace preservation of $\mathcal{L}$ implies that $\tr\mathcal{M}(\rho)=m_0\tr\rho=m_0$, where $\mathcal{P}_0$  is the projection for the zero eigenvalue of $\mathcal{L}$. Therefore, this quantity is trivial. This is related to the fact that Noether's theorem breaks down for Lindbladian systems~\cite{Victor}, and to the fact that we are talking about a superoperator structure on top of the usual observable space. Very recently, however, Stylaris and Zanardi showed~\cite{Zanardi} that for each conserved superoperator $\mathcal{M}$ satisfying $[\mathcal{M},\mathcal{L}]=0$ one can define a monotone function
\begin{equation}
f_{\mathcal{M}}(\rho)=\tr[\mathcal{M}(\rho)^\dagger(\mathbf{L}_\rho +\lambda \mathbf{R}_\rho )^{-1} (\mathcal{M}(\rho))], 
\end{equation}
with $\lambda \ge 0$, where $\mathbf{L}_\rho (X) = \rho X$ and $\mathbf{R}_\rho (X) = X\rho$ are the superoperators of left and right multiplication by $\rho$, respectively, and the inverse is well defined for strictly positive $\rho$.
 They showed that such a monotone, as complicated as it might look at first glance, is well motivated from entropic distances, and \emph{is  decreasing} under the evolution  $\rme^{t\mathcal{L}}$:
\begin{equation}
f_{\mathcal{M}}(\rho_t) \leq f_{\mathcal{M}}(\rho), \quad \text{for all } t\ge0,
\end{equation}
where $\rho_t = \rme^{ t\mathcal{L}} \rho$.

\begin{figure}
\includegraphics[width=0.48\textwidth]{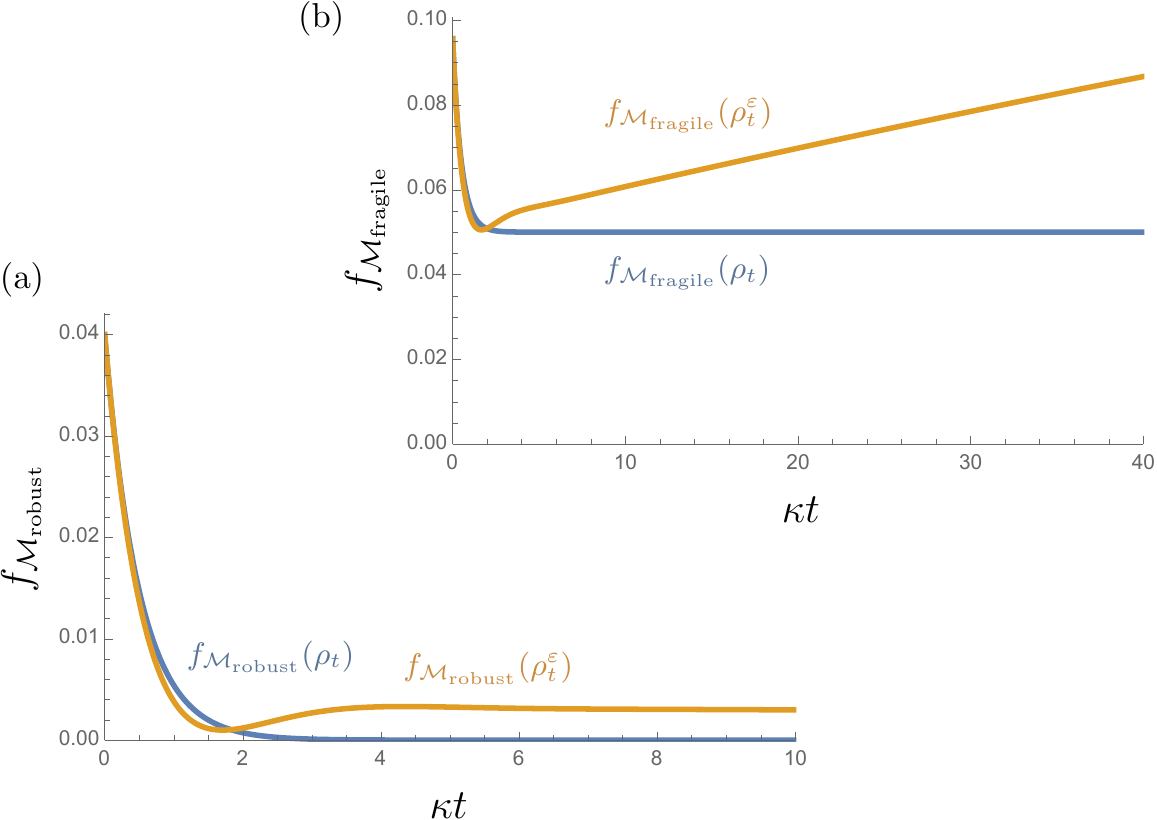}
\caption{The evolutions of (a) the robust monotone $f_{\mathcal{M}_\mathrm{robust}}(\rho)=|\bra{0}\rho\ket{1}|^2$ defined with $\mathcal{M}_\mathrm{robust}=-\frac{\rmi}{\sqrt{2}}[\sigma_z,{}\cdot{}\,]$ and $\lambda=1$, and of (b) the fragile monotone $f_{\mathcal{M}_\mathrm{fragile}}(\rho)$ defined with $\mathcal{M}_\mathrm{fragile}=-\frac{\rmi}{\sqrt{2}}[\sigma_z,{}\cdot{}\,]+\ket{0}\bra{0}{}\cdot{}\ket{0}\bra{0}$ and $\lambda=1$, under the unperturbed qubit evolution by $\mathcal{L}=-\frac{\rmi}{2}\omega[\sigma_z,{}\cdot{}\,]-\frac{1}{2}\kappa(1-\sigma_z{}\cdot{}\,\sigma_z)$ and under the evolution perturbed by $\varepsilon\mathcal{V}=-\frac{\rmi}{2}\varepsilon g[\sigma_x,{}\cdot{}\,]$, which creates coherence between $\ket{0}$ and $\ket{1}$. The states $\ket{0}$ and $\ket{1}$ are the eigenstates of $\sigma_z$. The initial state of the qubit is given by the coherence vector $(r_x,r_y,r_z)=(0.2,0,0.8)$, and the parameters are set at $g=\kappa$ and $\varepsilon=0.1$ ($\omega$ is irrelevant).}
\label{fig:Monotones}
\end{figure}

Using our generalized eternal block-diagonal approximation $\mathcal{V}_{\mathcal{L}}(\varepsilon)$ to a perturbation $\mathcal{V}$ for open systems \cite{eternal}, we can write the perturbed dynamics
\begin{equation}
\rme^{t(\mathcal{L}+\varepsilon \mathcal{V})}= \rme^{t[\mathcal{L}+\varepsilon \mathcal{V}_{\mathcal{L}}(\varepsilon)]}+O(\varepsilon),
\end{equation}
for all times, and see that, for any \emph{robust} symmetry $\mathcal{M}=\sum_k m_k \mathcal{P}_k$ and for any perturbation $\varepsilon \mathcal{V}$, 
the monotone $f_\mathcal{M}(\rho)$ remains approximately monotonic \cite{note:KAM_SM},
\begin{equation}
f_{\mathcal{M}}(\rho^{\varepsilon}_t)\le f_{\mathcal{M}}(\rho) + O(\varepsilon) ,
\end{equation}
under the perturbed evolution $\rho_t^\varepsilon = \rme^{ t(\mathcal{L}+\varepsilon \mathcal{V})} \rho$.
In this sense, the monotone is robust against perturbation.
See Fig.~\ref{fig:Monotones} for a couple of examples for a qubit dephasing evolution.
The monotone defined with $\mathcal{M}_\mathrm{robust}=\sum_km_k\mathcal{P}_k$ is robust against perturbation: it is perturbed and becomes nonmonotonic, but the nonmonotonicity is small.
On the other hand, the monotone defined with $\mathcal{M}_\mathrm{fragile}$ that lifts the degeneracy in $\mathcal{L}$ is fragile. 

\emph{Conclusions.---}%
While our results in spirit reproduce a lot of features one would hope a quantum KAM theory to feature---long-term stability of certain observables with respect to perturbations, in analogy with the KAM theory in classical mechanics~\cite{ref:ThirringClassicalPhysics,ref:ArnoldBook}---there are also some perhaps surprising aspects. Conserved charges and generalized Gibbs states from quantum integrable models are not robust, while randomly chosen Hamiltonians (thus without degeneracies) have the property that all conserved quantities are robust.

\begin{acknowledgments}
This research was funded in part by the Australian Research Council (project number FT190100106), and by the Top Global University Project from the Ministry of Education, Culture, Sports, Science and Technology (MEXT), Japan.
PF and SP were partially supported by Istituto Nazionale di Fisica Nucleare (INFN) through the project ``QUANTUM''. PF and SP acknowledge support by MIUR via PRIN 2017 (Progetto di Ricerca di Interesse Nazionale), project QUSHIP (2017SRNBRK). PF was partially supported by the Italian National Group of Mathematical Physics (GNFM-INdAM). PF and SP were partially supported by Regione Puglia and by QuantERA ERA-NET Cofund in Quantum Technologies (GA No.\ 731473), project PACE-IN\@.
HN is partly supported by the Institute for Advanced Theoretical and Experimental Physics, Waseda University and by Waseda University Grant for Special Research Projects (Project Number: 2020C-272).
KY was supported by the Grants-in-Aid for Scientific Research (C) (No.~18K03470) and for Fostering Joint International Research (B) (No.~18KK0073) both from the Japan Society for the Promotion of Science (JSPS).
\end{acknowledgments}

\begin{widetext}
\section{Supplemental Material}
In this Supplemental Material, we discuss the mathematical details.

\subsection{Zeno dynamics. Error bound}
Consider the spectral resolution of $H=H^\dagger$:
\begin{equation}
H=\sum_{k=1}^d h_k P_k,
\end{equation}
where
\begin{equation}
P_k P_\ell =\delta_{k\ell}P_\ell = \delta_{k\ell}P_\ell^\dagger, \qquad \sum_k P_k = \openone,
\end{equation}
$d\leq \dim \mathcal{H}$ is the number of distinct eigenvalues of $H$, and $h_k\in \mathbb{R}$, with $h_k\neq h_{\ell}$ for $k\neq \ell$.
Given a perturbation $V=V^\dagger$ its diagonal part (Zeno Hamiltonian) is given by
\begin{equation}
V_{\mathrm{Z}}=  \sum_k P_k V P_k.
\end{equation}
We want to bound the divergence 
\begin{equation}
\delta_{\mathrm{Z}}(t) = \|\rme^{\rmi t (H+\varepsilon V)} - \rme^{\rmi t (H+\varepsilon V_{\mathrm{Z}})} \|
\end{equation}
between the dynamics generated by $H+\varepsilon V$ and the  dynamics generated by its block-diagonal part $H+\varepsilon V_{\mathrm{Z}}$.
We will use a trick elaborated in Ref.~\cite{ref:unity1}, which is based on Kato's seminal proof of the adiabatic theorem~\cite{ref:KatoAdiabatic}.

Fix a spectral projection $P_\ell$ and consider the reduced resolvent at $h_\ell$, $\lim_{z\to h_\ell} (H-z\openone)^{-1} (\openone - P_\ell)$, that is
\begin{equation}
S_\ell = \sum_{k \,:\, k\neq \ell} \frac{1}{h_k - h_\ell} P_k.
\end{equation}
In the following, we will use $1$ for the identity operator $\openone$ and simply write $H-z$ instead of $H-z\openone$. 
We get $P_\ell S_\ell = S_\ell P_\ell = 0$ and
\begin{equation}
(H-h_\ell) S_\ell = S_\ell (H-h_\ell) = \sum_{k \,:\, k\neq \ell} \frac{h_k - h_\ell}{h_k-h_\ell} P_k = \sum_{k \,:\, k\neq \ell} P_k = 1- P_\ell,
\label{eq:invres}
\end{equation}
that is $S_\ell$ is the inverse of $H-h_\ell$ on the subspace range of $1-P_\ell$. 
We get
\begin{equation}
\rme^{\rmi t (H+\varepsilon V)} - \rme^{\rmi t (H+\varepsilon V_{\mathrm{Z}})} =
-\int_0^t \d s\,\frac{\partial}{\partial s} \bigl( \rme^{\rmi (t-s)(H+\varepsilon V)} \rme^{\rmi s(H+\varepsilon V_{\mathrm{Z}}) }\bigr)
= \rmi\varepsilon  \int_0^t \d s \, \rme^{\rmi (t-s)(H+\varepsilon V)} (V-V_{\mathrm{Z}})\rme^{\rmi s(H+\varepsilon V_{\mathrm{Z}}) },
\end{equation}
whence
\begin{equation}
\bigl(\rme^{\rmi t (H+\varepsilon V)} - \rme^{\rmi t (H+\varepsilon V_{\mathrm{Z}})} \bigr) P_\ell
= \rmi\varepsilon  \int_0^t \d s\,  \rme^{\rmi (t-s)(H+\varepsilon V)} (1 - P_\ell) V P_\ell \rme^{\rmi s(h_\ell+\varepsilon V_{\mathrm{Z}}) }.
\end{equation}
By~\eqref{eq:invres} we have
\begin{equation}
\bigl(\rme^{\rmi t (H+\varepsilon V)} - \rme^{\rmi t (H+\varepsilon V_{\mathrm{Z}})} \bigr) P_\ell
= \rmi\varepsilon  \int_0^t \d s \, \rme^{\rmi (t-s)(H+\varepsilon V)} (H- h_\ell) S_\ell V P_\ell \rme^{\rmi s(h_\ell+\varepsilon V_{\mathrm{Z}}) }.
\end{equation}
Now notice that
\begin{equation}
\rmi \rme^{\rmi (t-s)(H+\varepsilon V)} (H- h_\ell) = - \frac{\partial}{\partial s} \bigl( \rme^{\rmi (t-s) (H+\varepsilon V)} \rme^{\rmi s(h_\ell + \varepsilon V)}
\bigr)  \rme^{-\rmi s(h_\ell + \varepsilon V)},
\end{equation}
and thus
\begin{equation}
\bigl(\rme^{\rmi t (H+\varepsilon V)} - \rme^{\rmi t (H+\varepsilon V_{\mathrm{Z}})} \bigr) P_\ell
= - \varepsilon  \int_0^t \d s \, 
\frac{\partial}{\partial s} \bigl( \rme^{\rmi (t-s) (H+\varepsilon V)} \rme^{\rmi s(h_\ell + \varepsilon V)}
\bigr)  \rme^{-\rmi s \varepsilon V}
S_\ell V P_\ell \rme^{\rmi s \varepsilon V_{\mathrm{Z}} }.
\end{equation}
By integrating by parts
\begin{align}
\bigl(\rme^{\rmi t (H+\varepsilon V)} - \rme^{\rmi t (H+\varepsilon V_{\mathrm{Z}})} \bigr) P_\ell
={} & {-\varepsilon}  \int_0^t \d s  \,
\frac{\partial}{\partial s} \bigl( \rme^{\rmi (t-s) (H+\varepsilon V)} \rme^{\rmi s(h_\ell + \varepsilon V)}
  \rme^{-\rmi s \varepsilon V}
S_\ell V P_\ell \rme^{\rmi s \varepsilon V_{\mathrm{Z}} } \bigr) 
\nonumber\\
& {} + \varepsilon  \int_0^t \d s  \,
\rme^{\rmi (t-s) (H+\varepsilon V)} \rme^{\rmi s(h_\ell + \varepsilon V)}
  \frac{\partial}{\partial s} \bigl(  \rme^{-\rmi s \varepsilon V}
S_\ell V P_\ell \rme^{\rmi s \varepsilon V_{\mathrm{Z}} } \bigr) 
\nonumber\\
 ={} &
 \varepsilon  \bigl( \rme^{\rmi t (H+\varepsilon V)} S_\ell V P_\ell
- S_\ell V P_\ell \rme^{\rmi t (H +\varepsilon V_{\mathrm{Z}}) }  \bigr) 
\nonumber\\
& {}-\rmi \varepsilon^2  \int_0^t \d s  \,
\rme^{\rmi (t-s) (H+\varepsilon V)} 
(V S_\ell V P_\ell - S_\ell V P_\ell V_{\mathrm{Z}}) 
\rme^{\rmi s (H+ \varepsilon V_{\mathrm{Z}} )}  .
\end{align}
Finally, by summing over $\ell$ we have
\begin{equation}
\rme^{\rmi t (H+\varepsilon V)} - \rme^{\rmi t (H+\varepsilon V_{\mathrm{Z}})} 
= \varepsilon  \bigl( \rme^{\rmi t (H+\varepsilon V)} X
- X \rme^{\rmi t (H +\varepsilon V_{\mathrm{Z}}) }  \bigr) 
 -\rmi \varepsilon^2  \int_0^t \d s  \,
\rme^{\rmi (t-s) (H+\varepsilon V)} 
(V X - X V_{\mathrm{Z}}) 
\rme^{\rmi s (H+ \varepsilon V_{\mathrm{Z}} )}  ,
\end{equation}
where 
\begin{equation}
	X= \sum_\ell S_\ell V P_\ell.
	\label{eq:Xdef}
\end{equation}
By taking the operator norm, one gets
\begin{equation}
\delta_{\mathrm{Z}}(t) = \|\rme^{\rmi t (H+\varepsilon V)} - \rme^{\rmi t (H+\varepsilon V_{\mathrm{Z}})} \|
\leq 2 \varepsilon \|X\| +  \varepsilon^2  \int_0^t \d s  \,
(\|V\| \|X\|  +\| X \|  \|V_{\mathrm{Z}}\| ) =  2 \varepsilon \|X\| +  \varepsilon^2  \|X\| 
(\|V\| + \| V_{\mathrm{Z}}\| ) t.
\label{eq:bounddelta}
\end{equation}
Now, we get
\begin{equation}
\| X \|^2 = \| X X^\dag\| = \left\| 
\sum_\ell S_\ell V P_\ell V S_\ell 
\right\| \leq  \sum_\ell  \| S_\ell V P_\ell V S_\ell\|
\leq
\sum_\ell \|S_\ell\|^2 \|V\|^2,
\end{equation}
while
\begin{equation}
\|S_\ell\| =\left\| 
\sum_{k \,:\, k\neq \ell} \frac{P_k}{h_k-h_\ell} 
\right\| 
= \max_{k \,:\, k\neq \ell}  \left|  
\frac{1}{h_k-h_\ell}
\right|
\leq \frac{1}{\eta},
\end{equation}
where 
\begin{equation}
\eta = \min_{k,\ell \,:\, k\neq\ell} |h_k - h_\ell |
\end{equation}
is the minimum spectral gap of $H$,
and thus
\begin{equation}
\| X \| \leq  \frac{\sqrt{d}}{\eta} \|V\|.
\label{eq:boundX}
\end{equation}
Moreover, in the operator norm,
\begin{equation}
\|V_{\mathrm{Z}}\| = \left\| 
\sum_k P_k V P_k 
\right\| = \max_k \|P_k V P_k \| \leq \|V\|.
\label{eq:boundXVZ}
\end{equation}
Therefore, by plugging~\eqref{eq:boundX} and~\eqref{eq:boundXVZ} into~\eqref{eq:bounddelta},
we finally get
\begin{equation}
\delta_{\mathrm{Z}}(t) = \|\rme^{\rmi t (H+\varepsilon V)} - \rme^{\rmi t (H+\varepsilon V_{\mathrm{Z}})} \|
 \leq\frac{2 \sqrt{d}}{\eta}\varepsilon\|V\|(1+\varepsilon\|V\|t),
 \label{eq:bounddelta1}
\end{equation}
which for $\|V\|=1$ reduces to Eq.~(\ref{eq:deltaZ}) of the Letter.

\subsection{Robust symmetries}
Consider now a robust symmetry
\begin{equation}
M = \sum_km_kP_k,
\end{equation}
with $m_k\in\mathbb{R}$.
This is a conserved observable, $M=M^\dagger$, $[M, H]=0$, that acts uniformly within each eigenspace of $H$.
We have $M_t = \rme^{\rmi t H } M  \rme^{-\rmi t H } = M$, and for every perturbation $\varepsilon V$,
\begin{align}
\|M_t^{\varepsilon} - M\| 
&= \| \rme^{\rmi t (H +\varepsilon V)}  M \rme^{-\rmi t (H +\varepsilon V)}  -M \|
\nonumber\\
&=\| \rme^{\rmi t (H +\varepsilon V)}  M   -M \rme^{\rmi t (H +\varepsilon V)} \|
\nonumber\\
& = \bigl\| \bigl(\rme^{\rmi t (H +\varepsilon V)}  -  \rme^{\rmi t (H +\varepsilon V_{\mathrm{Z}})} \bigr) M   +  \rme^{\rmi t (H +\varepsilon V_{\mathrm{Z}})}  M 
-M \rme^{\rmi t (H +\varepsilon V)} \bigr\|.
\end{align}
By making use of the commutativity $[M, V_{\mathrm{Z}}] =0$, one gets
\begin{equation}
\|M_t^{\varepsilon} - M\| 
= \bigl\| \bigl(\rme^{\rmi t (H +\varepsilon V)}  -  \rme^{\rmi t (H +\varepsilon V_{\mathrm{Z}})} \bigr) M   
-M \bigl(\rme^{\rmi t (H +\varepsilon V)}  -  \rme^{\rmi t (H +\varepsilon V_{\mathrm{Z}})} \bigr)  \bigr\|
\leq 2 \| M\|   \| \rme^{\rmi t (H +\varepsilon V)}  -  \rme^{\rmi t (H +\varepsilon V_{\mathrm{Z}})}  \| ,
\end{equation}
that is
\begin{equation}
\|M_t^{\varepsilon} - M\| \leq 2 \| M\| \delta_{\mathrm{Z}}(t),
\end{equation}
which is the inequality~(4) of the Letter.

Analogously, by substituting in the previous derivation $V_{\mathrm{Z}}$ with $V_H(\varepsilon)$, which still commutes with the robust conserved observable $M$, i.e.\ $[M, V_{H}(\varepsilon)] =0$, one has the bound
\begin{equation}
\|M_t^{\varepsilon} - M\| 
= \bigl\| \bigl(\rme^{\rmi t (H +\varepsilon V)}  -  \rme^{\rmi t[H +\varepsilon V_{H}(\varepsilon)]}\bigr) M
-M \bigl(\rme^{\rmi t (H +\varepsilon V)}  -  \rme^{\rmi t[H +\varepsilon V_{H}(\varepsilon)]} \bigr)  \bigr\|
\leq 2 \| M\| \delta_\infty,
\label{eq:Minf}
\end{equation}
where 
\begin{equation}
\delta_\infty = \sup_t   \| \rme^{\rmi t (H +\varepsilon V)}  -  \rme^{\rmi t[H +\varepsilon V_{H}(\varepsilon)]} \|
\label{eq:25}
\end{equation}
is the uniform bound on the divergence of the two dynamics. 
This is the first inequality in Eq.~(9) of the Letter.

The block-diagonal perturbation $V_{H}(\varepsilon)$ can be chosen such that $\delta_\infty = O(\varepsilon)$.
The crucial ingredient is to choose a block-diagonal perturbation $H+\varepsilon V_H(\varepsilon)$ which is \emph{isospectral}  with $H+\varepsilon V$, and  thus is unitarily equivalent to it: 
\begin{equation}
H+ \varepsilon V_H(\varepsilon) = W_\varepsilon^\dagger (H+\varepsilon V) W_\varepsilon, 
\label{eq:isospec0}
\end{equation} 
with a unitary $W_\varepsilon = 1 + O(\varepsilon)$. 
Such a block-diagonal $V_H(\varepsilon)$ and a unitary $W_\varepsilon$ actually exist \cite{Cloizeaux,Klein,eternal}.
By plugging~\eqref{eq:isospec0} into~\eqref{eq:25}, we get
\begin{align}
\delta_\infty & 
 = \sup_t\| \rme^{\rmi t(H+\varepsilon V)} - W_\varepsilon^\dagger  \rme^{\rmi t (H+ \varepsilon V)} W_\varepsilon \| 
 \nonumber\\
 & = \sup_t\| (1 - W_\varepsilon^\dagger) \rme^{\rmi t(H+\varepsilon V)} + W_\varepsilon^\dagger  \rme^{\rmi t (H+ \varepsilon V)} (1 - W_\varepsilon) \|
 \nonumber\\ 
 & \leq \sup_t\| 1 - W_\varepsilon^\dagger \|  \| \rme^{\rmi t(H+\varepsilon V)} \| + \| W_\varepsilon^\dagger  \rme^{\rmi t (H+ \varepsilon V)}\|  \| 1 - W_\varepsilon \|
 \nonumber\\
 & = \| 1 - W_\varepsilon^\dagger \| +  \| 1 - W_\varepsilon \| 
\vphantom{\sup_t}
\nonumber\\
&= O(\varepsilon).
\label{eq:28}
\end{align}

The existence, and the explicit construction, of  a unitary $W_\varepsilon$, that carries the perturbed Hamiltonian into a block-diagonal form, is proved and discussed in detail in Ref.~\cite{eternal}. 
Here, in the next subsections, we will show the \emph{necessity} of an isospectral perturbation, and then discuss its construction and prove that
\begin{equation}
V_H(\varepsilon)  = V_{\mathrm{Z}} + O(\varepsilon),
\end{equation}
by exploiting the connection with  quantum KAM theory.

\subsection{Isospectral perturbations}
Consider a Hamiltonian $H=H^\dagger$ and a perturbation $\tilde{H} = H +O(\varepsilon)$, with small $\varepsilon$. 
We want to compare the two dynamics by looking at their divergence:
\begin{equation}
\delta_{H, \tilde{H}} (t) = \|\rme^{\rmi t \tilde{H}}- \rme^{\rmi t {H}}\| .
\end{equation}
Consider the spectral decompositions
\begin{equation}
H= \sum_{k=1}^d h_k P_k , \qquad \tilde{H} = \sum_{k=1}^d \tilde{h}_k \tilde{P}_k, 
\end{equation}
where $d$ is the number of distinct eigenvalues of $\tilde{H}$, i.e.\ $\tilde{h}_k\neq\tilde{h}_\ell$ for $k\neq\ell$.
It may happen that $h_k=h_\ell$ for some $k\neq\ell$, if the degeneracy is lifted by the perturbation. 
However in such a case we choose the orthogonal projections $P_k$ and $P_\ell$ such that they are adapted to the perturbation, that is $\tilde{P}_k = P_k + O(\varepsilon)$  and $\tilde{P}_\ell = P_\ell + O(\varepsilon)$~\cite{ref:KatoBook}.
As for the eigenvalues, $\tilde{h}_k = h_k + O(\varepsilon)$.

We get
\begin{equation}
\rme^{\rmi t \tilde{H}}- \rme^{\rmi t {H}} = \sum_k \bigl( \rme^{\rmi t \tilde{h}_k } \tilde{P}_k - \rme^{\rmi t h_k} P_k \bigr)
=  \sum_k  \rme^{\rmi t \tilde{h}_k } (\tilde{P}_k - P_k) 
- \sum_k (\rme^{\rmi t \tilde{h}_k } -\rme^{\rmi t h_k}) P_k .
\end{equation}
The first sum on the right-hand side is $O(\varepsilon)$ uniformly in time, as
\begin{equation}
\left\| 
\sum_k  \rme^{\rmi t \tilde{h}_k } (\tilde{P}_k - P_k)
\right\| \leq \sum_k \| \tilde{P}_k - P_k \| = O(\varepsilon).
\end{equation}
On the other hand, the last term reads 
\begin{equation}
\left\| \sum_k (\rme^{\rmi t \tilde{h}_k } -\rme^{\rmi t h_k}) P_k \right\| = \max_k |\rme^{\rmi t \tilde{h}_k} -\rme^{\rmi t h_k}| = 2 \max_k \left| \sin\!\left(t \frac{\tilde{h}_k - h_k}{2}\right)\right|,
\end{equation}
so that
\begin{equation}
\delta_{H,\tilde{H}} (t) = 2 \max_k \left| \sin  \left(t \frac{\tilde{h}_k - h_k}{2}\right)\right| + O(\varepsilon).
\label{eq:deltaHH}
\end{equation}
Therefore, since  $\tilde{h}_k - h_k =  O(\varepsilon)$, we get
\begin{equation}
\delta_{H,\tilde{H}}(t) = O(\varepsilon), \qquad \text{for} \quad t=O(1).
\end{equation}
However, the divergence has a slow drift (secular term) and becomes $O(1)$ for sufficiently large times $O(1/\varepsilon)$. 
Indeed,
\begin{equation}
\delta_{H, \tilde{H}} (t) = 2 + O(\varepsilon), \qquad \text{for} \quad t=\frac{\pi} {\tilde{h}_k - h_k} = O(1/\varepsilon),
\end{equation}
that is, the maximal divergence
\begin{equation}
\delta_\infty = \sup_t \delta_{H,\tilde{H}} (t) = 2 + O(\varepsilon).
\end{equation}

Geometrically, the evolution of a Hamiltonian with $d$ distinct eigenvalues yields a (quasi-)periodic motion of a point on a torus. 
Two motions with different frequencies, however small the differences may be, will eventually accumulate a divergence of $O(1)$.
The only way to avoid this slow drift is that the two motions be isochronous, that is the first term in~\eqref{eq:deltaHH} should be identically zero. 
This means that
\begin{equation}
\delta_\infty = O(\varepsilon) \qquad \text{iff} \qquad \tilde{h}_k = h_k, \qquad \text{for all } k ,
\end{equation}
i.e., the Hamiltonian $H$ and its perturbation $\tilde{H}$ must be isospectral.

\subsection{Quantum KAM iteration. Homological equation}
We are looking for a unitary transformation $W_\varepsilon$ close to the identity, such that the transformed total Hamiltonian is isospectral to $H+\varepsilon V$,
\begin{equation}
H+\varepsilon V_H(\varepsilon) = W_\varepsilon^\dagger (H+\varepsilon V) W_\varepsilon,
\label{eq:isospec}
\end{equation}
with the constraint that $V_H(\varepsilon)$ be block-diagonal,
\begin{equation}
V_H = \langle V_H \rangle := \sum_k P_k V_H P_k.
\label{eq:diag}
\end{equation}
By writing 
\begin{equation}
W_\varepsilon = \rme^{\rmi  K(\varepsilon)}, \qquad
K(\varepsilon) = \varepsilon K_1  + O(\varepsilon^2),
\end{equation}
with $K_1 = K_1^\dagger$, and
\begin{equation}
  V_H(\varepsilon) = V_0 +  O(\varepsilon),
\end{equation}
with $V_0=V_0^\dagger$,
Eq.~\eqref{eq:isospec} reads
\begin{equation}
H+ \varepsilon V_H(\varepsilon) =  (1  - \rmi \varepsilon K_1) (H+\varepsilon V)  (1  + \rmi \varepsilon K_1) + O(\varepsilon^2),
\end{equation}
whence
\begin{equation}
V_0 =   \rmi  [H, K_1]  +  V.
\label{eq:step1}
\end{equation}
Notice that 
\begin{equation}
\langle [H, K_1] \rangle = \sum_k P_k ( H K_1- K_1 H) P_k = \sum_k P_k (h_k K_1  - K_1 h_k ) P_k = 0.
\end{equation}
Therefore, the constraint~\eqref{eq:diag}, which implies $V_0 = \langle V_0 \rangle$, gives
\begin{equation}
V_0 = \langle V \rangle  = \sum_k P_k V P_k = V_{\mathrm{Z}},
\end{equation}
and
\begin{equation}
 \rmi [H, K_1] =- \{V\}, 
\label{eq:homological}
\end{equation}
where 
\begin{equation}
\{V\}:= V- \langle V \rangle =  \sum_{k,\ell \,:\, k\neq \ell} P_k V P_\ell = \sum_k P_k V (1 - P_k) = \frac{1}{2} \sum_k [P_k,[P_k,V]]
\end{equation}
is the off-diagonal part of $V$.

The expression~\eqref{eq:homological} should be understood as an equation for $K_1$, the first-order term of the generator $K(\varepsilon)$ of the unitary $W_\varepsilon$. It is known as the \emph{homological} equation and is the fundamental block of quantum KAM theory~\cite{ref:Sinai,ref:Russman,ref:Craig,ref:Poschel,ref:Bellissard}. It is the quantum analog of the  homological equation of KAM theory in classical mechanics, where the commutator is replaced by ($-\rmi$ times) the Poisson bracket, while $\langle{}\cdot{}\rangle$ and $\{{}\cdot{}\}$ are replaced by the averaged and the oscillating part of the perturbation, respectively~\cite{ref:ThirringClassicalPhysics,ref:ArnoldBook}.

One can prove that the homological equation~\eqref{eq:homological} has a unique solution with $\langle K_1 \rangle=0$, for every $H$ and $V$. Indeed, by  sandwiching~\eqref{eq:homological} between $P_k$ and $P_\ell$ with $k\neq \ell$ we get
\begin{equation}
(h_k - h_\ell) P_k K_1 P_\ell = \rmi  P_k V P_\ell,
\end{equation}
that is
\begin{equation}
\{ K_1\} = \rmi  \sum_{k,\ell \,:\, k\neq \ell} \frac{P_k V P_\ell}{h_k-h_\ell} = \rmi \sum_\ell S_\ell V P_\ell.
\label{eq:solutionh}
\end{equation}

Notice that $\{K_1\} = \{K_1\}^\dagger$, as it should be, and in fact one has
\begin{equation}
\{ K_1 \} = \rmi \sum_\ell S_\ell V P_\ell = -\rmi \sum_\ell P_\ell V S_\ell = \frac{\rmi}{2} \sum_\ell (S_\ell V P_\ell - P_\ell V S_\ell).
\end{equation}
Moreover, notice that we have a complete freedom in the choice of the block-diagonal part $\langle K_1 \rangle$ of $K_1$, since it commutes with $H$ and thus is immaterial in equation~\eqref{eq:homological}, so that
\begin{equation}
K_1 = \rmi \sum_\ell S_\ell V P_\ell  +   \sum_\ell P_\ell Z P_\ell,
\end{equation}
with an arbitrary $Z=Z^\dagger$. In the following, for simplicity, we will  \emph{fix the gauge} $Z=0$, i.e.\ $\langle K_1 \rangle=0$, and thus  will make the solution of~\eqref{eq:homological} unique.

From the explicit expression of the generator $K_1$, we can now easily evaluate a uniform bound on the divergence~\eqref{eq:25}. From the inequality~\eqref{eq:28}, we get
\begin{equation}
	\delta_\infty = \sup_t   \| \rme^{\rmi t (H +\varepsilon V)}  -  \rme^{\rmi t [H +\varepsilon V_{H}(\varepsilon)]} \| \leq 2  \| 1 - W_\varepsilon \| \leq 2\varepsilon \|K_1\| +O(\varepsilon^2) \leq \frac{2 \sqrt{d}}{\eta}\varepsilon\|V\| + O(\varepsilon^2),
\label{eq:firstorderbound}
\end{equation}
where the last inequality is a consequence of the bound~\eqref{eq:boundX}, since $K_1=\rmi X$. 

In fact, an explicit bound on the divergence $\delta_\infty$ is obtained in Ref.~\cite[Appendix~E]{eternal} as
\begin{equation}
\delta_\infty
\le \hat{\delta}_\infty, \qquad \text{where } \quad
\hat{\delta}_\infty = 2\sqrt{d}\left(
\frac{1}{\sqrt[4]{1-4\varepsilon/\eta}}-1
\right) = \frac{2 \sqrt{d}}{\eta}\varepsilon + O(\varepsilon^2),
\end{equation}
for $\|V\|=1$, which is easily seen to be always larger than the first order term in~\eqref{eq:firstorderbound},
$\hat{\delta}_\infty  \geq 2 \sqrt{d} \,\varepsilon/ \eta$.

This bound becomes trivial once it exceeds $\delta_\infty=2$ as $\varepsilon$ increases.
Since $d\ge2$, let us care only about the values of $\varepsilon$ where $2\sqrt{2}\,(1/\sqrt[4]{1-4\varepsilon/\eta}-1)\le2$, namely, for $4\varepsilon/\eta\le(13 + 12\sqrt{2})/(17 + 12 \sqrt{2})= x_0$.
Within this range, one gets the linear bound  $2(1/\sqrt[4]{1-4\varepsilon/\eta}-1)\le (\sqrt{2}/ x_0) 4\varepsilon/\eta 
< 7 \varepsilon/\eta$. 
Therefore, we have
\begin{equation}
\delta_\infty <  \frac{7\sqrt{d}}{\eta} \varepsilon.
\end{equation}
This yields Eq.~(7) of the Letter.

\subsubsection{Higher-order terms}
One can also show that all the following steps of the KAM iteration, giving higher-order terms $V_n$ in $V_H(\varepsilon)$ and $K_{n+1}$ in $K(\varepsilon)$, with $n\geq 1$, have the same structure as the first step and involve homological equations. For example, by considering the next-order terms,
\begin{equation}
K(\varepsilon) = \varepsilon K_1 + \varepsilon^2 K_2 +  O(\varepsilon^3) , \qquad  V_H(\varepsilon) = V_0 +  \varepsilon V_1 + O(\varepsilon^2),
\end{equation}
one gets
\begin{equation}
H+ \varepsilon V_0 + \varepsilon^2 V_1 
= (H+\varepsilon V) +\rmi \varepsilon [H+\varepsilon V, K_1] - \frac{1}{2} \varepsilon^2 [[H, K_1] , K_1] +\rmi \varepsilon^2 [H, K_2] + O(\varepsilon^3).
\end{equation}
The second-order terms give
\begin{equation}
V_1 =  \rmi  [H,K_2] -\frac{1}{2} [[H, K_1], K_1] +  \rmi  [V,K_1] ,
\label{eq:step2}
\end{equation}
that is
\begin{equation}
V_1 = \rmi  [H,K_2] + \rmi \left[V- \frac{1}{2}\{ V\}, K_1\right]. 
 \end{equation}
This has the same structure as~\eqref{eq:step1}, and gives
\begin{equation}
V_1 = \langle V_1\rangle = \left\langle \rmi \left[V- \frac{1}{2}\{ V\}, K_1\right] \right\rangle = - \sum_\ell    P_\ell V  S_\ell V P_\ell,
\end{equation}
and a homological equation for $K_2$:
\begin{equation}
 \rmi  [H,K_2]  = - \left\{ \rmi \left[V- \frac{1}{2}\{ V\}, K_1\right]  \right\}.
 \end{equation}

In general, at order $\varepsilon^{n+1}$ one gets an equation of the form
\begin{equation}
V_n = \rmi [H, K_{n+1}] + P_n(\mathcal{K}_1, \dots, \mathcal{K}_n)(H) + Q_n(\mathcal{K}_1, \dots, \mathcal{K}_n)(V),
\end{equation}
where $P_n$ and $Q_n$ are polynomials of order $n$ and $\mathcal{K}_j$ are the superoperators $\mathcal{K}_j(Y)= \rmi [Y, K_j]$.
This has the same structure as~\eqref{eq:step1} or~\eqref{eq:step2}.
$V_n$ will be given by the block-diagonal part of the right-hand side, while $K_{n+1}$ will be the solution of the homological equation given by the off-diagonal part.

This is the algebraic structure of the KAM iteration scheme. And for our purposes this is enough. 
See for example~\cite{ref:Scherer95,ref:Scherer97}.
However, most difficulties and the hardest part of this scheme arises for infinite-dimensional systems with a vanishing minimal spectral gap $\eta$ because of an accumulation point of the discrete spectrum. Interesting cases are  systems with dense point spectrum~\cite{ref:Sinai,ref:Russman,ref:Craig,ref:Poschel,ref:Bellissard}. In such a situation, at each iteration step, the  solution of the homological equation~\eqref{eq:solutionh} suffers from the plague of \emph{small denominators}, the same problem that besets celestial mechanics.
 The reduced resolvent $S_\ell$ becomes unbounded, and the formal expression~\eqref{eq:solutionh} is a bounded operator only for a particular class of perturbations $V$ which are adapted to the Hamiltonian $H$: the closer are the eigenvalues $h_k$ and $h_\ell$ of $H$ at the denominator of~\eqref{eq:solutionh}, the smaller must be the numerator $P_k V P_\ell$. In such a case, the proof of the existence and the convergence of the series  makes use of classical techniques of KAM perturbation theory with a careful control of small denominators through a Diophantine condition, and a super-convergent iteration scheme~\cite{ref:ThirringClassicalPhysics,ref:ArnoldBook}.

\subsection{Robustness of monotones}
In Ref.~\cite{Zanardi}, it is shown that for a symmetry $\mathcal{M}$ of a Lindbladian $\mathcal{L}$ satisfying $[\mathcal{M},\mathcal{L}]=0$ one can define a monotone
\begin{equation}
f_{\mathcal{M}}(\rho)=\tr[\mathcal{M}(\rho)^\dagger(\mathbf{L}_\rho +\lambda \mathbf{R}_\rho )^{-1} (\mathcal{M}(\rho))],
\end{equation}
which decreases under the evolution $\rho_t=\rme^{t\mathcal{L}}\rho$,
\begin{equation}
f_{\mathcal{M}}(\rho_t) \leq f_{\mathcal{M}}(\rho), \quad \text{for all } t\ge0,
\end{equation}
where $\mathbf{L}_\rho (X) = \rho X$ and $\mathbf{R}_\rho (X) = X\rho$ are the superoperators of left and right multiplication by $\rho$, respectively, and the inverse with $\lambda \ge 0$ is well defined for strictly positive $\rho$.
Here, we prove that a monotone defined with respect to a symmetry of the form
\begin{equation}
\mathcal{M}=\sum_km_k\mathcal{P}_k,
\label{eqn:RobustMopen}
\end{equation}
where $\{\mathcal{P}_k\}$ are the spectral projections of the Lindbladian $\mathcal{L}$, remains a monotone up to an error $O(\varepsilon)$ eternally even in the presence of a perturbation $\varepsilon\mathcal{V}$, namely,
\begin{equation}
f_{\mathcal{M}}(\rho_t^\varepsilon) \leq f_{\mathcal{M}}(\rho)+O(\varepsilon), \quad \text{for all } t\ge0,
\label{eqn:PerturbedMonotonicity}
\end{equation}
where $\rho_t^\varepsilon=\rme^{t(\mathcal{L}+\varepsilon\mathcal{V})}\rho$.
In this sense, $\mathcal{M}$ in (\ref{eqn:RobustMopen}) is a robust symmetry of the evolution $\mathcal{L}$.

To show this, we first note that even in the case of open-system evolution one can find a block-diagonal approximation $\mathcal{V}_\mathcal{L}(\varepsilon)$ of the perturbation $\mathcal{V}$ such that $\mathcal{L}+\varepsilon\mathcal{V}_\mathcal{L}(\varepsilon)$ is similar to $\mathcal{L}+\varepsilon\mathcal{V}$ \cite{eternal},
\begin{equation}
\mathcal{L}+\varepsilon\mathcal{V}_\mathcal{L}(\varepsilon)
=\mathcal{W}_\varepsilon^{-1}(\mathcal{L}+\varepsilon\mathcal{V})\mathcal{W}_\varepsilon.
\end{equation}
Then, let us consider 
\begin{equation}
\tilde{\mathcal{M}}=\mathcal{W}_\varepsilon\mathcal{M}\mathcal{W}_\varepsilon^{-1}.
\end{equation}
This is a symmetry of the perturbed system $\mathcal{L}+\varepsilon\mathcal{V}$, corresponding to the symmetry $\mathcal{M}$ of the unperturbed system $\mathcal{L}$, since $[\mathcal{M},\mathcal{V}_\mathcal{L}(\varepsilon)]=0$.
Since this similarity transformation is small, $\mathcal{W}_\varepsilon=1+O(\varepsilon)$, we have 
\begin{equation}
\tilde{\mathcal{M}}=\mathcal{M}+O(\varepsilon).
\end{equation}
Notice that the monotone $f_{\tilde{\mathcal{M}}}(\rho)$ defined with respect to $\tilde{\mathcal{M}}$ is decreasing under the perturbed evolution $\rho_t^\varepsilon=\rme^{t(\mathcal{L}+\varepsilon\mathcal{V})}\rho$.
Therefore,
\begin{equation}
f_{\mathcal{M}}(\rho_t^\varepsilon)
=f_{\tilde{\mathcal{M}}}(\rho_t^\varepsilon)+O(\varepsilon)
\le f_{\tilde{\mathcal{M}}}(\rho)+O(\varepsilon)
=f_\mathcal{M}(\rho)+O(\varepsilon), \quad \text{for all } t\ge0.
\end{equation}
This proves the approximate monotonicity (\ref{eqn:PerturbedMonotonicity}).
\end{widetext}


\begin{thebibliography}{10}

\bibitem{Gross}
D.~J. Gross, The Role of Symmetry in Fundamental Physics, \href{https://doi.org/10.1073/pnas.93.25.14256}{Proc. Natl. Acad. Sci. USA \textbf{93}, 14256 (1996)}.

\bibitem{Moretti}
V. Moretti, \href{https://doi.org/10.1007/978-3-030-18346-2}{\textit{Fundamental Mathematical Structures of Quantum Theory}} (Springer, Cham, 2019).

\bibitem{Alhashimi}
M. Al-Hashimi, \href{http://www.wiese.itp.unibe.ch/theses/al-hashimi_phd.pdf}{\textit{Accidental Symmetry in Quantum Physics}}, Ph.D. Thesis, Institute for Theoretical Physics, Bern University, Bern, 2008.

\bibitem{Caux}
G. P. Brandino, J.-S. Caux, and R. M Konik, Glimmers of a Quantum KAM Theorem: Insights from Quantum Quenches in One-Dimensional Bose Gases, \href{https://doi.org/10.1103/PhysRevX.5.041043}{Phys. Rev. X \textbf{5}, 041043 (2015).}

\bibitem{Nori}
I.~M. Georgescu, S. Ashhab, and F. Nori, Quantum Simulation, \href{https://doi.org/10.1103/RevModPhys.86.153}{Rev. Mod. Phys. \textbf{86}, 153 (2014)}.

\bibitem{Sarovar}
M. Sarovar, J. Zhang, and L. Zeng, Reliability of Analog Quantum Simulation, \href{https://doi.org/10.1140/epjqt/s40507-016-0054-4}{EPJ Quantum Technol. \textbf{4}, 1 (2017)}.

\bibitem{Schwenk}
I. Schwenk, J.-M. Reiner, S. Zanker, L. Tian, J. Lepp\"akangas, and M. Marthaler, Reconstructing the Ideal Results of a Perturbed Analog Quantum Simulator, \href{https://doi.org/10.1103/PhysRevA.97.042310}{Phys. Rev. A \textbf{97}, 042310 (2018)}.

\bibitem{ref:ThirringClassicalPhysics}
W. Thirring, \href{https://doi.org/10.1007/978-1-4612-0681-1}{\textit{Classical Mathematical Physics: Dynamical Systems and Field Theories}, 3rd ed.} (Springer, New York, 1997).

\bibitem{ref:ArnoldBook}
V.~I. Arnold, V.~V. Kozlov, and A.~I. Neishtadt, \href{https://doi.org/10.1007/978-3-540-48926-9}{\textit{Mathematical Aspects of Classical and Celestial Mechanics}, 3rd ed.} (Springer, Berlin, 2006).

\bibitem{ref:Scherer95}
W. Scherer, Superconvergent Perturbation Method in Quantum Mechanics, \href{https://doi.org/10.1103/PhysRevLett.74.1495}{Phys. Rev. Lett. \textbf{74}, 1495 (1995)}.

\bibitem{ref:QZS}
P. Facchi and S. Pascazio, Quantum {Zeno} Subspaces, \href{https://doi.org/10.1103/PhysRevLett.89.080401}{Phys. Rev. Lett. \textbf{89}, 080401 (2002)}.

\bibitem{ref:PaoloSaverio-QZEreview-JPA}
P. Facchi and S. Pascazio, Quantum {Zeno} Dynamics: Mathematical and Physical Aspects, \href{https://doi.org/10.1088/1751-8113/41/49/493001}{J. Phys. A: Math. Theor. \textbf{41}, 493001 (2008)}.

\bibitem{ref:unity1}
D. Burgarth, P. Facchi, H. Nakazato, S. Pascazio, and K. Yuasa, Generalized Adiabatic Theorem and Strong-Coupling Limits, \href{https://doi.org/10.22331/q-2019-06-12-152}{Quantum \textbf{3}, 152 (2019)}.

\bibitem{HamazakiPRL2020}
Z. Gong, N. Yoshioka, N. Shibata, and R. Hamazaki, Universal Error Bound for Constrained Quantum Dynamics, \href{https://doi.org/10.1103/PhysRevLett.124.210606}{Phys. Rev. Lett. \textbf{124}, 210606 (2020)}.

\bibitem{note:KAM_SM}
See the Supplemental Material.

\bibitem{eternal}
D. Burgarth, P. Facchi, H. Nakazato, S. Pascazio, and K. Yuasa, Eternal Adiabaticity, \href{https://arxiv.org/abs/2011.04713}{arXiv:2011.04713 [quant-ph]}.

\bibitem{ref:MatrixFunctions-Higham}
N.~J. Higham, \href{https://doi.org/10.1137/1.9780898717778}{\textit{Functions of Matrices: Theory and Computation}} (SIAM, Philadelphia, 2008).

\bibitem{ref:Caux-QuantumIntegrability}
J.-S. Caux and J. Mossel, Remarks on the Notion of Quantum Integrability, \href{https://doi.org/10.1088/1742-5468/2011/02/p02023}{J. Stat. Mech. \textbf{2011}, P02023 (2011)}.

\bibitem{ref:Braak-RabiExact}
D. Braak, Integrability of the Rabi Model, \href{https://doi.org/10.1103/PhysRevLett.107.100401}{Phys. Rev. Lett. \textbf{107}, 100401 (2011)}.

\bibitem{Grabowski}
M.~P. Grabowski and P. Mathieu, Quantum Integrals of Motion for the {Heisenberg} Spin Chain, \href{https://doi.org/10.1142/S0217732394002057}{Mod. Phys. Lett. A \textbf{09}, 2197 (1994)}.

\bibitem{Haag}
R. Haag, D. Kastler, and E.~B. Trych-Pohlmeyer, Stability and Equilibrium States, \href{https://doi.org/10.1007/BF01651541}{Commun. Math. Phys. \textbf{38}, 173 (1974)}.

\bibitem{ref:PolkovnikovRMP}
A. Polkovnikov, K. Sengupta, A. Silva, and M. Vengalattore, Nonequilibrium Dynamics of Closed Interacting Quantum Systems, \href{https://link.aps.org/doi/10.1103/RevModPhys.83.863}{Rev. Mod. Phys. \textbf{83}, 863 (2011)}.

\bibitem{ref:EisertFriesdorfGogolin-ReviewNatPhys}
J. Eisert, M. Friesdorf, and C. Gogolin, Quantum Many-Body Systems Out of Equilibrium, \href{https://doi.org/10.1038/nphys3215}{Nat. Phys. \textbf{11}, 124 (2015)}.

\bibitem{ref:GooldReviewJPA}
J. Goold, M. Huber, A. Riera, L. {del Rio}, and P. Skrzypczyk, The Role of Quantum Information in Thermodynamics: A Topical Review, \href{https://doi.org/10.1088/1751-8113/49/14/143001}{J. Phys. A: Math. Theor. \textbf{49}, 143001 (2016)}.

\bibitem{ref:GogolinEisertReview}
C. Gogolin and J. Eisert, Equilibration, Thermalisation, and the Emergence of Statistical Mechanics in Closed Quantum Systems, \href{https://doi.org/10.1088/0034-4885/79/5/056001}{Rep. Prog. Phys. \textbf{79},  056001 (2016)}.

\bibitem{ref:VidmarRigolGGEreview}
L. Vidmar and M. Rigol, Generalized {Gibbs} Ensemble in Integrable Lattice Models, \href{https://doi.org/10.1088/1742-5468/2016/06/064007}{J. Stat. Mech. \textbf{2016}, 064007 (2016)}.

\bibitem{ref:CalabresePhysicaA}
P. Calabrese, Entanglement and Thermodynamics in Non-Equilibrium Isolated Quantum Systems, \href{https://doi.org/10.1016/j.physa.2017.10.011}{Physica A \textbf{504}, 31 (2018)}.

\bibitem{ref:MoriIkedaKaminishiUeda-ReviewJPB}
T. Mori, T.~N. Ikeda, E. Kaminishi, and M. Ueda, Thermalization and Prethermalization in Isolated Quantum Systems: A Theoretical Overview, \href{https://doi.org/10.1088/1361-6455/aabcdf}{J. Phys. B: At. Mol. Opt. Phys. \textbf{51}, 112001 (2018)}.

\bibitem{Victor}
V.~V. Albert and L. Jiang, Symmetries and Conserved Quantities in {Lindblad} Master Equations, \href{https://doi.org/10.1103/PhysRevA.89.022118}{Phys. Rev. A \textbf{89}, 022118 (2014)}.

\bibitem{Zanardi}
G. Styliaris and P. Zanardi, Symmetries and Monotones in {Markovian} Quantum Dynamics, \href{https://doi.org/10.22331/q-2020-04-30-261}{Quantum \textbf{4}, 261 (2020)}.




\bibitem{ref:KatoAdiabatic}
T. Kato, On the Adiabatic Theorem of Quantum Mechanics, \href{https://doi.org/10.1143/JPSJ.5.435}{J. Phys. Soc. Jpn. \textbf{5}, 435 (1950)}.

\bibitem{Cloizeaux}
J. {des Cloizeaux}, Extension d'une Formule de Lagrange \`a des Probl\`emes de Valeurs Propres, \href{https://doi.org/10.1016/0029-5582(60)90177-2}{Nucl. Phys. \textbf{20}, 321 (1960)}.

\bibitem{Klein}
D.~J. Klein, Degenerate Perturbation Theory, \href{https://doi.org/10.1063/1.1682018}{J. Chem. Phys. \textbf{61}, 786 (1974)}.


\bibitem{ref:KatoBook}
T. Kato, \href{https://doi.org/10.1007/978-3-642-66282-9}{\textit{Perturbation Theory for Linear Operators}, 2nd ed.} (Springer, Berlin, 1976).

\bibitem{ref:Sinai}
E.~I. Dinaburg and Y.~G. Sinai, The One-Dimensional {Schr\"odinger} Equation with a Quasiperiodic Potential, \href{https://doi.org/10.1007/BF01075873}{Funct. Anal. Its Appl. \textbf{9}, 279 (1975)}.

\bibitem{ref:Russman}
H. R\"ussmann, On the One-Dimensional {Schr\"odinger} Equation with a Quasi-Periodic Potential, \href{https://doi.org/10.1111/j.1749-6632.1980.tb29679.x}{Ann. NY Acad. Sci. \textbf{357}, 90 (1980)}.

\bibitem{ref:Craig}
W. Craig, Pure Point Spectrum for Discrete Almost Periodic {Schr\"odinger} Operators, \href{https://doi.org/10.1007/BF01206883}{Commun. Math. Phys. \textbf{88}, 113 (1983)}.

\bibitem{ref:Poschel}
J. P\"oschel, Examples of Discrete {Schr\"odinger} Operators with Pure Point Spectrum, \href{https://doi.org/10.1007/BF01211953}{Commun. Math. Phys. \textbf{88}, 447 (1983)}.

\bibitem{ref:Bellissard}
J. Bellissard, R. Lima, and E. Scoppola, Localization in $\nu$-Dimensional Incommensurate Structures, \href{https://doi.org/10.1007/BF01211954}{Commun. Math. Phys. \textbf{88}, 465 (1983)}.




\bibitem{ref:Scherer97}
W. Scherer, Quantum Averaging {II}: {Kolmogorov's} Algorithm, \href{https://doi.org/10.1088/0305-4470/30/8/026}{J. Phys. A: Math. Gen. \textbf{30}, 2825 (1997)}.


\end{thebibliography}
\end{document}